# More considerations about the symmetry of the stress tensor of fluids

Ji Luo[*]

*College of Optoelectronic Engineering, Chengdu University of Information Technology,*

*Chengdu 610225, Sichuan, China*

**ABSTRACT**

Regarding a recent dispute about the symmetry of the stress tensor of fluids, more considerations are presented. The usual proofs of this symmetry are reviewed, and contradictions between this symmetry and the mechanism of gas viscosity are analyzed for simple gas flows. It is emphasized that these proofs may not be valid because they depend on the theorem of angular momentum which preassumes that the internal forces between any two fluid particles are always along the line connecting them. From Newton's laws of motion, however, one can only obtain the theorem of angular momentum with an additional term. It is proved that this additional term represents the total moment of internal forces in a fluid, both by using continuum model and by considering the microscopic structure of the fluid. In the latter case, this term has the form of the total moment of the forces exerted on the nuclei by the electrons. This moment of internal forces may lead to nonsymmetry of the stress tensor and is in general nonzero as long as shear stress exists. A nonsymmetrical stress tensor suggested in the literature is discussed in terms of its effect in eliminating the contradictions and simplifying the Navier-Stokes equation. The derivation of this stress tensor for ideal gases based on the kinetic theory of gas molecules is presented, and its form in a general orthogonal curvilinear coordinate system is given. Finally, a possible experimental verification of this nonsymmetrical stress tensor is discussed.

## 1. Introduction

The symmetry of the stress tensor of ordinary fluids is a widely accepted conclusion and its derivations can be found in usual textbooks. Although there have been researches on fluids with nonsymmetrical stress tensors since a very early time [1-6], these researches were mainly related to unusual cases such as fluids subject to external volume or surface couples [1], or fluids involving macromolecules [4,6]. Text books usually state the absence of concentrated coupled stresses [7], or mention the possible nonsymmetry of the stress tensor for the fluid having a local torque (moment) proportional to the volume in external field [8,9]. Recently, this symmetry was questioned by [10] in a general sense, based on the analysis of inconsistencies in specific flows where the stress tensor is taken as symmetrical. The article also questioned the validity of the conservation of angular momentum for fluids which is crucial to the derivation of the symmetry of the stress tensor, and suggested a simpler nonsymmetrical stress tensor which eliminates the inconsistencies and at the same time keeps the Navier-Stokes equation unchanged under a certain condition. This work was later criticized by another work [11] which defends the existing textbook conclusions. In the present work, we further consider the nonsymmetry of the stress tensor with regard to ordinary fluids, that is, our work does not concern fluids subject to couples especially, nor does it stress fluids with macromolecules in particular.

---

[*] Email address: jluo@cuit.edu.cn

Although the inconsistencies discussed in [10] were refuted by [11], obvious contradictions still exist in the simplest laminar flow and the uniform rotation of a gas, between the symmetrical stress tensor and the mechanism of gas viscosity. According to the mechanism, there should be no viscous force along one direction perpendicular to the laminar flow, but the symmetrical stress tensor gives the same force as that parallel to the flow. Similarly, the mechanism exists in the uniform rotation of the gas, but the symmetrical stress tensor leads to zero viscous force. These contradictions indicate that the derivation of the symmetry of the stress tensor may need to be reexamined, in terms of its theoretical basis and assumptions. Probably some important factors were missed or misunderstood. On the other hand, the symmetry of the stress tensor is usually regarded as a firm conclusion of the continuum model of fluids, its correctness being undoubted [12]. It thus may be enlightening to examine possible limitations of the continuum model. After all, this model is a mathematical abstraction. In practice, the microscopic structure of the fluid may play a role [2-5], and one may need to stop at some level in dividing the material, due to the lack of knowledge of the deeper levels. Such a stop seems possible because structures and motions of deeper levels may not affect the fluid motion. All these indicate that the doubt about the symmetry of the stress tensor for ordinary fluids is reasonable, and at least gases may have a nonsymmetrical stress tensor. In view of the contradictions and limitations, it seems that the solution of the dispute about the symmetry may finally depend on experiment.

The above issues will be discussed in this work. In Section two we review the derivation of the symmetry of the stress tensor adopted in text books and present contradictions involved in real gas flows. In section three we discuss a preassumption in the existing derivation and analyze its validity. In section four, we discuss a symmetrical stress tensor suggested in [10]. In section five, a possible experimental verification of the symmetry or nonsymmetry is suggested. Finally we present the conclusions in section six.

## 2. The existing derivation and contradictions
### 2.1. The existing derivation

Suppose in a rectangular coordinate system, $\vec{p}_x = p_{xx}\vec{i} + p_{xy}\vec{j} + p_{xz}\vec{k}$, $\vec{p}_y = p_{yx}\vec{i} + p_{yy}\vec{j} + p_{yz}\vec{k}$, $\vec{p}_z = p_{zx}\vec{i} + p_{zy}\vec{j} + p_{zz}\vec{k}$ are the forces respectively exerted on three unit areas in a fluid each perpendicular to a coordinate axis. The components, $p_{xx}$, $p_{xy}$, etc., constitute the stress tensor $p_{xx}\vec{i}\vec{i} + p_{xy}\vec{i}\vec{j} + \cdots$ and we adopt the convention that the first subscript of a component denotes normal direction of the area, and the second indicates direction of the component [13]. The force $\vec{p}_n = p_{nx}\vec{i} + p_{ny}\vec{j} + p_{nz}\vec{k}$ exerted on a unit area having the unit normal $\vec{n} = n_x\vec{i} + n_y\vec{j} + n_z\vec{k}$ can be expressed as $\vec{p}_n = n_x\vec{p}_x + n_y\vec{p}_y + n_z\vec{p}_z$. The equation of motion of the fluid is then [8]

$$\rho \frac{d\vec{v}}{dt} = \rho\vec{f} + \frac{\partial \vec{p}_x}{\partial x} + \frac{\partial \vec{p}_y}{\partial y} + \frac{\partial \vec{p}_z}{\partial z}, \tag{1}$$

where $\rho$ is the density of the fluid, $\vec{v}$ is the velocity of a fluid particle, $\vec{f}$ is the density of external force, $d/dt = \partial/\partial t + \vec{v} \cdot \nabla$ is the total derivative and $\nabla = \vec{i}\partial/\partial x + \vec{j}\partial/\partial y + \vec{k}\partial/\partial z$.

The stress tensor is generally believed to be symmetrical, that is,

$$p_{xy} = p_{yx}, \quad p_{yz} = p_{zy}, \quad p_{zx} = p_{xz}. \tag{2}$$

The usual proofs are based on the theorem of angular momentum, that is, for a volume $V$ of the fluid,

$$\frac{d}{dt}\iiint_V \vec{r} \times \rho\vec{v}\,dV = \iiint_V \vec{r} \times \rho\vec{f}\,dV + \oiint_S \vec{r} \times \vec{p}_n\,dS, \tag{3}$$

where $\vec{r}$ is the position vector of a fluid particle, $S$ is the surface of volume $V$, $n$ denotes the outward normal direction of $S$, and $d/dt$ denotes the substantial derivative. One starts from Eq. (3) and transforms the surface integral $\oiint_S \vec{r} \times \vec{p}_n\,dS$ into a volume one. Then according to the equation of motion of the fluid (1) one infers that the quantity

$$\vec{M} = \iiint_V [(p_{zy} - p_{yz})\vec{i} + (p_{xz} - p_{zx})\vec{j} + (p_{yx} - p_{xy})\vec{k}]\,dV \tag{4}$$

must be zero. Since the volume $V$ is arbitrarily chosen in the fluid, one concludes that the integrand in Eq. (4) vanishes, and the stress tensor is symmetrical.

*2.2. Two examples contradicting the symmetrical stress tensor*

Nevertheless, there are real gas flows which contradict the symmetry of the stress tensor, if the mechanism of the gas viscosity is noted. First consider the simplest horizontal laminar flow discussed in [10] but now for a gas [7,8], that is, the velocity $\vec{v} = ay\vec{i}$ with $\vec{i}$ being horizontal, $\vec{j}$ being vertical, and $a$ being a positive constant. According to the symmetrical stress tensor, since $v_x = ay$, $v_y = 0$, $v_z = 0$, the fluid has the shear stress $p_{yx} = \mu(\partial v_x/\partial y + \partial v_y/\partial x) = \mu a$ where $\mu$ is the coefficient of viscosity. The mechanism of this viscosity of the gas is explained like this [14]. Because $v_x = ay$ increases with $y$, the gas molecules above a horizontal area (perpendicular to $\vec{j}$) have larger horizontal momentum than those below the area. Thus as molecules cross the area due to thermal motion, horizontal momentum is transported from the upper part of the gas to the lower one, and a horizontal viscous force is accordingly generated whose magnitude can be calculated according to the amount of horizontal momentum transported. The symmetrical stress tensor requires that the gas also has the shear stress $p_{xy} = p_{yx} = \mu a$. Physically, however, it seems unreasonable that a vertical area (perpendicular to $\vec{i}$) will be acted upon by a vertical force in the direction of $\vec{j}$, since gas molecules on the two sides of this area have the same zero vertical momentum, and the mechanism of momentum transportation is absent in this case. Experiment confirmed the stress $p_{yx} = \mu a$ and the viscosity law, but the stress $p_{xy}$ lacks experimental measurements for ordinary gases. It is thus probable that $p_{xy} = 0$ instead of $p_{xy} = p_{yx}$.

The second example is about the uniform rotation of fluids also mentioned in [10]. Suppose a portion of gas is contained in a cylinder and the system rotates around its axis which is taken as z-axis. In a cylindrical coordinate system having unit coordinate vectors $\vec{e}_r$, $\vec{e}_\theta$, $\vec{e}_z$, the constant angular velocity is $\vec{\omega} = \omega \vec{e}_z$. The gas particles then move with the tangential velocity $v_\theta = \omega r$, where $r$ is the radial distance of the particle from the z-axis. Since this velocity increases with $r$, there should be a transport of tangential momentum, in the direction of a radius due to the thermal motion of the molecules. As the result, there should be a tangential viscous force on an area perpendicular to the radius, that is, $p_{r\theta} \neq 0$. But from symmetrical stress tensor one has $p_{r\theta} = p_{\theta r} = \mu(\partial v_\theta / \partial r + r^{-1} \partial v_r / \partial \theta - v_\theta / r) = 0$. Again it seems that stress tensors of gases might be nonsymmetrical.

## 3. Reconsidering the symmetry of the stress tensor, the moment of internal forces
### 3.1. A preassumption in the existing derivation and the missing term

In proving the symmetry of the stress tensor, the application of the theorem of angular momentum in the form of Eq. (3) needs a condition to be satisfied, that is, the moment of internal forces of the fluid disappears [11,12]. This may be made clear by briefly reviewing the derivation of Eq. (3). As a conclusion of the continuum model, Eq. (3) originates from the same theorem for a system of mass particles, that is,

$$\frac{d}{dt}\sum_{i=1}^{N} \vec{r}_i \times m_i \vec{v}_i = \sum_{i=1}^{N} \vec{r}_i \times \vec{f}_i + \sum_{i=1}^{N}\sum_{j \neq i} \vec{r}_i \times \vec{f}_{i,j}, \qquad (5)$$

where $m_i$, $\vec{r}_i$, $\vec{v}_i$, $\vec{f}_i$ are respectively the mass, position vector, velocity, and external force of the $i$th particle, and $\vec{f}_{i,j}$ is the internal force exerted on the $i$th particle by the $j$th particle. (The indexes $i$, $j$, $k$ should be distinguished form coordinate vectors $\vec{i}$, $\vec{j}$, $\vec{k}$.) A volume of continuum fluid is divided into small volumes and each small volume is approximately regarded as a mass particle, and Eq. (5) is applied to the small volumes with the assumption that the internal force for each pair of small volumes $\vec{f}_{i,j}$ is along the line connecting their positions $\vec{r}_i$ and $\vec{r}_j$ so that $\vec{r}_i \times \vec{f}_{i,j} + \vec{r}_j \times \vec{f}_{j,i} = (\vec{r}_i - \vec{r}_j) \times \vec{f}_{i,j} = 0$ for all $i \neq j$ due to Newton's third law. Hence for small volumes the last term in Eq. (5) vanishes. By a limiting procedure in which the sizes of all small volumes tend to zero, Eq. (5) leads to Eq. (3).

However, if one starts from the equation of motion (1) and does not assume the symmetry of the stress tensor, by taking the vector product of $\vec{r}$ and Eq. (1), integrating the result in $V$, and then transforming the surface integral into a volume one, one obtains

$$\frac{d}{dt}\iiint_V \vec{r} \times \rho\vec{v}dV = \iiint_V \vec{r} \times \rho\vec{f}dV + \oiint_S \vec{r} \times \vec{p}_n dS + \vec{M}, \tag{6}$$

with $\vec{M}$ expressed by Eq. (4). This means that from Newton's laws of motion, one can only obtain the theorem of angular momentum in the form of Eq. (6) instead of Eq. (3). The point is that Eq. (6) contains an additional term $\vec{M}$ which is not axiomatically zero.

*3.2. Investigating the additional term within the continuum model*

Equation (6) can be obtained through an integration of Eq. (2) in [1] where the term $\vec{M}$ is regarded as the angular momentum consumed or produced inside the fluid. We demonstrate here that in general $\vec{M}$ is the total moment of internal forces, and this conclusion is also obtained by using Eq. (5). Similar to the derivation of Eq. (3) for continuum fluids, one may use three sets of coordinate planes $x = x_i$, $y = y_j$, $z = z_k$ to divide the volume $V$ into small cuboids (some on the boundary of $V$ may have curved surfaces, and again the indexes $i$, $j$, $k$ should be distinguished form $\vec{i}$, $\vec{j}$, $\vec{k}$). The planes respectively have intervals $\Delta x$, $\Delta y$, $\Delta z$ and the cuboids have equal volume $\Delta x \Delta y \Delta z$ (except those on the boundary). One first approximately regards the cuboids as point particles and applies Eq. (5) to them. Then one takes a limiting procedure so that the summations are transformed into integrals as the size of the cuboids tends to zero. In this process, the left term $(d/dt)\sum_{i=1}^N \vec{r}_i \times m_i \vec{v}_i$ of Eq. (5) becomes $(d/dt)\iiint_V \vec{r} \times \rho\vec{v}dV$. Similarly, the term $\sum_{i=1}^N \vec{r}_i \times \vec{f}_i$ on the right side of Eq. (5) becomes $\iiint_V \vec{r} \times \rho\vec{f}dV + \oiint_S \vec{r} \times \vec{p}_n dS$. To get the term $\sum_{i=1}^N \sum_{j \neq i} \vec{r}_i \times \vec{f}_{i,j}$ on the right side of Eq. (5), one notes that each $\vec{f}_{i,j}$ reduces to one of $\pm\vec{p}_x$, $\pm\vec{p}_y$, $\pm\vec{p}_z$ multiplied by the area of a facet of a cuboid and acts upon this facet. For instance, at the interface $x = x_i$ separating two neighboring cuboids located along the x-axis, two stresses $\vec{p}_x \Delta y \Delta z$ and $-\vec{p}_x \Delta y \Delta z$ are respectively exerted on the left and the right cuboids. Suppose $\vec{r}_1$ and $\vec{r}_2$ are respectively the centers of the two cuboids and thus $\vec{r}_1 - \vec{r}_2 = -\Delta x \vec{i}$. The sum of the moments of this pair of forces then is $\vec{r}_1 \times \vec{p}_x \Delta y \Delta z - \vec{r}_2 \times \vec{p}_x \Delta y \Delta z = -\vec{i} \times \vec{p}_x \Delta x \Delta y \Delta z$. This demonstrates that if the stresses $\vec{p}_x$ and $-\vec{p}_x$ are not along the x-axis, or the line connecting the centers of the two neighboring cuboids, the sum of their moments will not be zero. And this is the case if $\vec{p}_x$ has tangential components, that is, $p_{xy} \neq 0$ or $p_{xz} \neq 0$. By adding all $-\vec{i} \times \vec{p}_x \Delta x \Delta y \Delta z$ together for $i$, $j$, $k$ and transforming the summation into an integral, one obtains the moment $-\vec{i} \times \iiint_V \vec{p}_x dV$. Along with similar treatment for the interfaces separating neighboring cuboids in y-direction and z-direction, one finds the total moment of internal forces $\sum_{i=1}^N \sum_{j \neq i} \vec{r}_i \times \vec{f}_{i,j}$ to be

$$-\vec{i} \times \iiint_V \vec{p}_x dV - \vec{j} \times \iiint_V \vec{p}_y dV - \vec{k} \times \iiint_V \vec{p}_z dV, \tag{7}$$

which is just $\vec{M} = \iiint_V [(p_{zy} - p_{yz})\vec{i} + (p_{xz} - p_{zx})\vec{j} + (p_{yx} - p_{xy})\vec{k}]dV$ in Eq. (4). Hence by the above derivation one obtains Eq. (6) again, but now $\vec{M}$ exhibits the physical meaning of the total moment of internal forces. Furthermore, the derivation demonstrates that as long as shear stress exists, then $\vec{M} \neq 0$ in general. Of course it is also probable that $\vec{M}$ as an integral is zero, although its integrand is not zero identically, but verification of this is needed.

### 3.3. Investigating the additional term microscopically

The continuum model of fluids is a mathematical abstraction, that is, the material structures of fluids are not considered and physical quantities are expressed with abstract derivatives or integrals. As the result, in the derivation of Eq. (3) from Eq. (5), it is assumed that whatever the size and shape of the small volume, the forces between any two small volumes are always along the line connecting them [11,12]. Nevertheless, real fluids have atomic or molecular structures, and the division of the fluid into small parts cannot be carried out endlessly. In practice, one needs to find a material level at which the constituent particles can be regarded as point particles, that is, their internal structures and internal motions are not related to the motion of the fluid. For these particles, the theorem of angular momentum in the form of Eq. (5) can be applied and the assumption about the directions of their interactions may be verified. At this level, the abstract integrals are substituted by summations which have real physical significance.

It is reasonable to assume that in usual cases, nuclei could be regarded as the above point particles and the theorem of angular momentum (5) can be applied to them. This is because first, the nucleus is much heavier than the electron and the motion of the fluid can thus be approximately represented by the collective motion of the nuclei if we neglect the mass of the electrons, and second, the nucleus is extreme small and its internal spin is usually not coupled with the flow of ordinary fluids, especially if external electromagnetic fields do not exist. As a charged particle, a nucleus is acted upon by other nuclei and by the cloud of its surrounding electrons, both through electromagnetic interactions. The interaction between any two nuclei can be regarded as along the line connecting them, and the corresponding terms $\vec{r}_i \times \vec{f}_{i,j}$ in Eq. (5) originating from these interactions thus are canceled. The interaction between a nucleus and the electronic cloud, however, seems rather complicated. Suppose the force exerted on the $i$ th nucleus by the electronic cloud is $\vec{f}_{i,e}$. The theorem of angular momentum for the fluid then reduces to

$$\frac{d}{dt}\sum_{i=1}^{N}\vec{r}_i \times m_i\vec{v}_i = \sum_{i=1}^{N}\vec{r}_i \times \vec{f}_i + \sum_{i=1}^{N}\vec{r}_i \times \vec{f}_{i,e}. \qquad (8)$$

where $m_i$, $\vec{r}_i$, $\vec{v}_i$, $\vec{f}_i$ are respectively the mass, position vector, velocity, and external force of the $i$ th nucleus. According to Eq. (8) we have

$$\vec{M} = \sum_{i=1}^{N}\vec{r}_i \times \vec{f}_{i,e}. \qquad (9)$$

This demonstrates from the microscopic point of view that $\vec{M}$ is the total moment of internal forces, as electrons are inside the fluid. It also reveals the origin of the moment of internal forces microscopically for ordinary fluids.

*3.4. About the balance of moments of forces on a fluid element*

Equation (6) holds for a small cuboid. Hence another usual proof of the symmetry of the stress tensor by applying the theorem of angular momentum (3) to a small cuboid and then letting its size tend to zero may also be invalid. In fact, if we apply Eq. (6) instead of Eq. (3) to the small cuboid with $\vec{r}$ starting from its center, the volume integrals $(d/dt)\iiint_V \vec{r} \times \rho\vec{v}\,dV = \iiint_V \vec{r} \times \rho(d\vec{v}/dt)\,dV$ and $\iiint_V \vec{r} \times \rho\vec{f}\,dV$ have zero main parts (they are higher order quantities than $\Delta x \Delta y \Delta z$), and the surface integral $\oiint_S \vec{r} \times \vec{p}_n\,dS$ has two times of $(\Delta x/2)\vec{i} \times \vec{p}_x \Delta y \Delta z + (\Delta y/2)\vec{j} \times \vec{p}_y \Delta z \Delta x + (\Delta z/2)\vec{k} \times \vec{p}_z \Delta x \Delta y$ as its main part which equals $[(p_{yz} - p_{zy})\vec{i} + (p_{zx} - p_{xz})\vec{j} + (p_{xy} - p_{yx})\vec{k}]\Delta x \Delta y \Delta z$. But there is the last volume integral $\vec{M} = \iiint_V [(p_{zy} - p_{yz})\vec{i} + (p_{xz} - p_{zx})\vec{j} + (p_{yx} - p_{xy})\vec{k}]\,dV$ whose main part is $[(p_{zy} - p_{yz})\vec{i} + (p_{xz} - p_{zx})\vec{j} + (p_{yx} - p_{xy})\vec{k}]\Delta x \Delta y \Delta z$. As the result, when applied to the cuboid, Eq. (6) in the limiting case leads to identity $0 = 0$ which gives no result. This is different from the usual case of the balance of moments of forces where Eq. (3) is applied and leads to $(p_{yz} - p_{zy})\vec{i} + (p_{zx} - p_{xz})\vec{j} + (p_{xy} - p_{yx})\vec{k} = 0$.

*3.5. Conclusions that can be deduced*

Hence in usual proofs of the symmetry of the stress tensor, the condition $\vec{M} = 0$ is preassumed so that Eq. (3) can be applied. Equation (3) holds if the moments of internal forces of any two fluid particles cancel each other, and this is true if the force and reacting force lie along the line connecting the two particles [12]. But this is not axiomatically true. One has to examine whether $\vec{M} = 0$ is satisfied.

According to Eq. (9), whether $\vec{M}$ is zero or not may depend ultimately on the distribution of the electronic cloud in the fluid. It is well known that electronic cloud counterpoises repulsions between nuclei and thus molecules are formed. Similarly, electronic cloud may also regulate forces between molecules and thus affects the motion of fluids. Complicated as it is, the summation in Eq. (9) may act as a starting point to investigate the moment of internal forces and its effect on the fluid flow.

## 4. About the nonsymmetrical stress tensor given in the literature

The effect of nonsymmetry of the stress tensor on the fluid motion is an interesting topic, since the equation of motion (1) involves only the combination of derivatives of the stress-tensor components and different stress tensors may lead to the same fluid motion. In [10], a nonsymmetrical stress tensor is suggested, which leads to the same Navier-Stokes equation as the symmetrical one either for incompressible flows or if the Stokes assumption about the relation between the two coefficients of viscosity is slightly modified. Here we neglect the second coefficient of viscosity since it is usually related to extreme conditions [14], and discuss the nonsymmetrical stress tensor given by

$$\vec{p}_x = -p\vec{i} + \mu \frac{\partial \vec{v}}{\partial x}, \quad \vec{p}_y = -p\vec{j} + \mu \frac{\partial \vec{v}}{\partial y}, \quad \vec{p}_z = -p\vec{k} + \mu \frac{\partial \vec{v}}{\partial z}, \tag{10}$$

where $p$ is the pressure and $\mu$ is the coefficient of viscosity. The constitutive equations (10) lead to

$$\rho \frac{d\vec{v}}{dt} = \rho \vec{f} - \nabla p + \mu \nabla^2 \vec{v}. \tag{11}$$

Formally, Eq. (11) is the Navier-Stokes equation for incompressible fluids, which is most usually investigated. However, here the incompressibility is not assumed for Eqs. (10,11). As the result, the constitutive equations (10) may have simplified the Navier-Stokes equation for compressible fluids. On the other hand, for compressible fluids especially for gases, $\mu$ may not be a constant, and an additional term $(\nabla \mu \cdot \nabla) \vec{v}$ may need to be added to the right side of Eq. (11).

The nonsymmetrical constitutive equations (10) can be derived for gases from the kinetic theory of gas molecules. A brief derivation of the viscous force based the momentum transportation is like this, which is a direct extension of the derivation for the simplest laminar flow to a general one [15]. One takes a small area $\Delta S$ in the gas at $P(x, y, z)$ which is perpendicular to the x-axis. According to the mechanism of momentum transportation, the viscous force exerted on the fluid on the $-x$ side of $\Delta S$ by that on the $+x$ side equals the increase of the momentum of the gas on the $-x$ side in a unit time interval, due to the net transport of momentum by the molecules crossing $\Delta S$. The number of molecules crossing $\Delta S$ in unit time along $-x$ direction and that along $+x$ direction are the same and supposed to be $n_0 \bar{v} \Delta S / 6$, with $n_0$ being the number density of the gas molecules at $P$ and $\bar{v}$ their average speed of thermal motion. Furthermore, the molecules crossing $\Delta S$ from $-x$ to $+x$ side are supposed to have the fluid velocity $\vec{v}(x - \lambda, y, z)$, and those crossing $\Delta S$ from $+x$ to $-x$ side, $\vec{v}(x + \lambda, y, z)$, where $\lambda$ is the mean free path of the molecules. As the result, the net transport of momentum along $-x$ direction is $n_0 \bar{v} \Delta S [m\vec{v}(x + \lambda, y, z) - m\vec{v}(x - \lambda, y, z)]/6$ with $m$ being the mass of the molecule. By employing the Taylor expansions $\vec{v}(x + \lambda, y, z) = \vec{v}(P) + \lambda \partial \vec{v}/\partial x$ and $\vec{v}(x - \lambda, y, z) = \vec{v}(P) - \lambda \partial \vec{v}/\partial x$ with $\partial \vec{v}/\partial x$ taking value at $P$ and by writing $\mu = n_0 m \bar{v} \lambda / 3$, one obtains the viscous force $\mu \Delta S \partial \vec{v}/\partial x$. Similar derivation for $\Delta S$ perpendicular to y-axis or z-axis leads to the viscous force $\mu \Delta S \partial \vec{v}/\partial y$ or $\mu \Delta S \partial \vec{v}/\partial z$. With the pressure $p$ included and by letting $\Delta S = 1$, one obtains Eqs. (10). This derivation demonstrates that even the simplest model of structureless gas molecules may lead to the nonsymmetrical stress tensor, though there are more complicated theories [2,16].

By performing a coordinate transformation on Eqs. (10), one obtains the nonsymmetrical stress-tensor components in cylindrical polar coordinates

$$p_{rr} = -p + \mu \frac{\partial v_r}{\partial r}, \quad p_{r\theta} = \mu \frac{\partial v_\theta}{\partial r}, \quad p_{rz} = \mu \frac{\partial v_z}{\partial r},$$

$$p_{\theta r} = \mu\left(\frac{1}{r}\frac{\partial v_r}{\partial \theta} - \frac{v_\theta}{r}\right), \quad p_{\theta\theta} = -p + \mu\left(\frac{1}{r}\frac{\partial v_\theta}{\partial \theta} + \frac{v_r}{r}\right), \quad p_{\theta z} = \frac{\mu}{r}\frac{\partial v_z}{\partial \theta}, \quad (12)$$

$$p_{zr} = \mu\frac{\partial v_r}{\partial z}, \quad p_{z\theta} = \mu\frac{\partial v_\theta}{\partial z}, \quad p_{zz} = -p + \mu\frac{\partial v_z}{\partial z}.$$

Similarly, a coordinate transformation of Eq. (1) without the assumption of the symmetry of the stress tensor leads to the equations of motion in the cylindrical coordinates

$$\rho\left(\frac{dv_r}{dt} - \frac{v_\theta^2}{r}\right) = \rho f_r + \frac{\partial p_{rr}}{\partial r} + \frac{1}{r}\frac{\partial p_{\theta r}}{\partial \theta} + \frac{\partial p_{zr}}{\partial z} + \frac{p_{rr}}{r} - \frac{p_{\theta\theta}}{r},$$

$$\rho\left(\frac{dv_\theta}{dt} + \frac{v_r v_\theta}{r}\right) = \rho f_\theta + \frac{\partial p_{r\theta}}{\partial r} + \frac{1}{r}\frac{\partial p_{\theta\theta}}{\partial \theta} + \frac{\partial p_{z\theta}}{\partial z} + \frac{p_{r\theta}}{r} + \frac{p_{\theta r}}{r}, \quad (13)$$

$$\rho\frac{dv_z}{dt} = \rho f_z + \frac{\partial p_{rz}}{\partial r} + \frac{1}{r}\frac{\partial p_{\theta z}}{\partial \theta} + \frac{\partial p_{zz}}{\partial z} + \frac{p_{rz}}{r},$$

where $d/dt = \partial/\partial t + v_r \partial/\partial r + r^{-1} v_\theta \partial/\partial \theta + v_z \partial/\partial z$. By substituting Eqs. (12) into Eqs. (13) one obtains the same equations of motion expressed in terms of $v_r$, $v_\theta$, $v_z$ (or Navier-Stokes equations) as those for the symmetrical stress tensor for incompressible fluids. More results for general orthogonal curvilinear coordinates are given in the Appendix. It seems that as long as the nonsymmetrical stress tensor leads to the same Navier-Stokes equation as the symmetrical one in a usual rectangular coordinate system, it also does so in a general orthogonal curvilinear coordinate system.

Contrasted with symmetrical stress tensor, the nonsymmetrical one in Eqs. (10) may thus give the same fluid motion but different internal forces. For the two examples in Section 2, the nonsymmetrical stress tensor gives the reasonable results $p_{yx} = \mu a$ and $p_{xy} = 0$ for the laminar flow, and $p_{r\theta} = \mu\omega$ for the rotating gas in a cylinder. It also gives $p_{\theta r} = -\mu\omega$ for the latter. This is also reasonable because in this case one has $\partial \vec{v}/\partial \theta = -\omega r \vec{e}_r$. Since the velocity $\vec{v} = \omega r \vec{e}_\theta$ varies with $\theta$, there should be a transport of radial momentum along the direction of $\vec{e}_\theta$ (radial momentum appears as $\theta$ increases because $\partial \vec{v}/\partial \theta = -\omega r \vec{e}_r$). Here we note a difference between molecules of a fluid and those of a solid. The former has the motion of diffusion and may not be regarded as moving collectively in the manner of a rigid body, even though macroscopically the fluid is in a pure rotation (without deformation). The effect of diffusion is particularly manifest in gases so that the friction may exist due to the collisions of diffusing molecules, no matter whether the gas has the motion of deformation or not.

## 5. A possible experimental test for the nonsymmetry of the stress tensor

The solution of the dispute about the symmetry of the stress tensor may ultimately depend on experimental tests. For the incompressible fluid rotating in a cylinder, the nonsymmetrical stress tensor (12) leads to a nonzero resisting moment of the viscous force exerted on the cylinder, whereas the symmetrical stress tensor gives zero result [7]. But the measurement of the moment of the viscous force seems difficult, especially for gases. Instead, the steady rotation of an incompressible fluid between two coaxial cylinders may provide a feasible experimental test for us to judge the two stress tensors, as different flows are resulted in. Take the axis of the cylinders as the z-axis of the cylindrical coordinate system and suppose the radii and the angular velocities of the inner and outer cylinders are $R_1$, $R_2$, $\omega_1$, $\omega_2$ respectively. The Navier-Stokes equation (11) gives the same expression of the velocity of the fluid for both the symmetrical and nonsymmetrical stress tensors [7]

$$v_\theta(r) = \frac{1}{R_2^2 - R_1^2}\left[(\omega_2 R_2^2 - \omega_1 R_1^2)r - \frac{(\omega_2 - \omega_1)R_1^2 R_2^2}{r}\right]. \tag{14}$$

The stress exerted on the inner cylinder is, however, different and can be expressed as

$$p_{r\theta}(R_1) = \mu\left[\frac{dv_\theta(R_1)}{dr} - \frac{v_\theta(R_1)}{R_1}\right], \quad p_{r\theta}(R_1) = \mu\frac{dv_\theta(R_1)}{dr} \tag{15}$$

respectively for the symmetrical stress tensor and for the nonsymmetrical one (see Eqs. (12)). Suppose the inner cylinder is acted upon by some other moment of friction $M_0 > 0$ except that of the viscous force of the fluid. For steady rotation the balance of the moments requires $2\pi R_1^2 L p_{r\theta}(R_1) = M_0$, with $L$ being the length of the inner cylinder. Together with Eqs. (14,15), this gives the different angular velocities of the inner cylinder for the symmetrical and nonsymmetrical stress tensors, which are respectively

$$\omega_1 = \omega_2 - \frac{R_2^2 - R_1^2}{4\pi\mu R_1^2 R_2^2 L}M_0, \quad \omega_1 = \frac{2R_2^2}{R_1^2 + R_2^2}\omega_2 - \frac{R_2^2 - R_1^2}{2\pi\mu(R_1^2 + R_2^2)R_1^2 L}M_0. \tag{16}$$

Since $R_1 < R_2$, one can infer that if the stress tensor is symmetrical, in steady rotation the angular velocity of the inner cylinder will never exceed that of the outer one ($\omega_1 \leq \omega_2$), whereas if the stress tensor is the nonsymmetrical one given in Eqs. (12), the inner cylinder will rotate faster than the outer one ($\omega_1 > \omega_2$) as long as $\omega_2 > M_0/2\pi\mu R_1^2 L$. This seems a practicable experiment as long as one can effectively diminish other moments of friction exerted on the inner cylinder except that of the viscous force of the fluid (a small moment of inertia of the inner cylinder may also facilitate the experiment).

Another interesting result of Eqs. (16) is that if the inner cylinder remains rotationless ($\omega_1 = 0$), then both the symmetrical and nonsymmetrical stress tensors give the same balancing moment of force $M_0 = 4\pi\mu R_1^2 R_2^2 L\omega_2/(R_2^2 - R_1^2)$ exerted on the inner cylinder. This is the quantity that a viscosimeter employing two coaxial cylinders measures in order to determine the viscosity coefficient $\mu$ of the fluid. Hence the nonsymmetry of the stress tensor will not affect results of the usual viscosity measurements with such a viscosimeter. Similar calculations on a sphere moving in a fluid, using spherical coordinates, demonstrate that the Stokes formula which gives the viscous force exerted on the sphere will also not be modified. As experimental measurements of stresses in fluids usually present considerable difficulties, theoretical determination may play a role.

## 6. Conclusions

In reality, there are simple gas flows in which the symmetrical stress tensor contradicts the mechanism of gas viscosity. The usual proofs of the symmetry of the stress tensor of fluids may not be valid, because they utilize a preassumption about the internal forces of the fluid. The general theorem of angular momentum derived from Newton's laws of motion contains an additional term which is not axiomatically zero. Within the continuum model of fluids, it can be proved by a usual calculus method that the additional term represents the total moment of internal forces. Microscopically, for ordinary fluids this term may have the form of the total moment of the forces exerted on the nuclei by the electrons, and more knowledge about the interactions between nuclei and electrons may provide us more insights into the stress tensor. A simple nonsymmetrical stress tensor given in the literature may eliminate the contradictions, and can be derived according to the kinetic theory of gas molecules. The same theory also explains these contradictions, in terms of molecular diffusion and collision. Besides, this nonsymmetrical stress tensor may lead to the simplest Navier-Stokes equation as that for incompressible fluids, both in a usual rectangular coordinate system and in a general orthogonal curvilinear coordinate system. Finally, the nonsymmetrical stress tensor may be verified by an experiment on the flow of fluid between two coaxial cylinders.

**Appendix**

For a system of orthogonal curvilinear coordinates $q_1$, $q_2$, $q_3$ with unit coordinate vectors $\vec{e}_1$, $\vec{e}_2$, $\vec{e}_3$, through a coordinate transformation the constitutive equations (10) become

$$\vec{p}_1 = \left[ -p + \frac{\mu}{h_1}\left( \frac{\partial v_1}{\partial q_1} + \frac{v_2}{h_2}\frac{\partial h_1}{\partial q_2} + \frac{v_3}{h_3}\frac{\partial h_1}{\partial q_3} \right) \right]\vec{e}_1 + \frac{\mu}{h_1}\left( \frac{\partial v_2}{\partial q_1} - \frac{v_1}{h_2}\frac{\partial h_1}{\partial q_2} \right)\vec{e}_2 + \frac{\mu}{h_1}\left( \frac{\partial v_3}{\partial q_1} - \frac{v_1}{h_3}\frac{\partial h_1}{\partial q_3} \right)\vec{e}_3,$$

$$\vec{p}_2 = \frac{\mu}{h_2}\left( \frac{\partial v_1}{\partial q_2} - \frac{v_2}{h_1}\frac{\partial h_2}{\partial q_1} \right)\vec{e}_1 + \left[ -p + \frac{\mu}{h_2}\left( \frac{\partial v_2}{\partial q_2} + \frac{v_3}{h_3}\frac{\partial h_2}{\partial q_3} + \frac{v_1}{h_1}\frac{\partial h_2}{\partial q_1} \right) \right]\vec{e}_2 + \frac{\mu}{h_2}\left( \frac{\partial v_3}{\partial q_2} - \frac{v_2}{h_3}\frac{\partial h_2}{\partial q_3} \right)\vec{e}_3,$$

$$\vec{p}_3 = \frac{\mu}{h_3}\left( \frac{\partial v_1}{\partial q_3} - \frac{v_3}{h_1}\frac{\partial h_3}{\partial q_1} \right)\vec{e}_1 + \frac{\mu}{h_3}\left( \frac{\partial v_2}{\partial q_3} - \frac{v_3}{h_2}\frac{\partial h_3}{\partial q_2} \right)\vec{e}_2 + \left[ -p + \frac{\mu}{h_3}\left( \frac{\partial v_3}{\partial q_3} + \frac{v_1}{h_1}\frac{\partial h_3}{\partial q_1} + \frac{v_2}{h_2}\frac{\partial h_3}{\partial q_2} \right) \right]\vec{e}_3,$$

where $h_1$, $h_2$, $h_3$ are Lamé's coefficients. The equation of motion (1) in this coordinate system becomes

$$\rho\left(\frac{dv_1}{dt}+\frac{v_1v_2}{h_1h_2}\frac{\partial h_1}{\partial q_2}+\frac{v_1v_3}{h_1h_3}\frac{\partial h_1}{\partial q_3}-\frac{v_2^2}{h_1h_2}\frac{\partial h_2}{\partial q_1}-\frac{v_3^2}{h_1h_3}\frac{\partial h_3}{\partial q_1}\right)=\rho f_1+\frac{1}{h_1h_2h_3}\left[\frac{\partial(h_2h_3 p_{11})}{\partial q_1}\right.$$

$$\left.+\frac{\partial(h_3h_1 p_{21})}{\partial q_2}+\frac{\partial(h_1h_2 p_{31})}{\partial q_3}\right]+\frac{p_{12}}{h_1h_2}\frac{\partial h_1}{\partial q_2}+\frac{p_{13}}{h_1h_3}\frac{\partial h_1}{\partial q_3}-\frac{p_{22}}{h_1h_2}\frac{\partial h_2}{\partial q_1}-\frac{p_{33}}{h_1h_3}\frac{\partial h_3}{\partial q_1},$$

$$\rho\left(\frac{dv_2}{dt}+\frac{v_2v_3}{h_2h_3}\frac{\partial h_2}{\partial q_3}+\frac{v_2v_1}{h_2h_1}\frac{\partial h_2}{\partial q_1}-\frac{v_3^2}{h_2h_3}\frac{\partial h_3}{\partial q_2}-\frac{v_1^2}{h_2h_1}\frac{\partial h_1}{\partial q_2}\right)=\rho f_2+\frac{1}{h_1h_2h_3}\left[\frac{\partial(h_2h_3 p_{12})}{\partial q_1}\right.$$

$$\left.+\frac{\partial(h_3h_1 p_{22})}{\partial q_2}+\frac{\partial(h_1h_2 p_{32})}{\partial q_3}\right]+\frac{p_{23}}{h_2h_3}\frac{\partial h_2}{\partial q_3}+\frac{p_{21}}{h_2h_1}\frac{\partial h_2}{\partial q_1}-\frac{p_{33}}{h_2h_3}\frac{\partial h_3}{\partial q_2}-\frac{p_{11}}{h_2h_1}\frac{\partial h_1}{\partial q_2},$$

$$\rho\left(\frac{dv_3}{dt}+\frac{v_3v_1}{h_3h_1}\frac{\partial h_3}{\partial q_1}+\frac{v_3v_2}{h_3h_2}\frac{\partial h_3}{\partial q_2}-\frac{v_1^2}{h_3h_1}\frac{\partial h_1}{\partial q_3}-\frac{v_2^2}{h_3h_2}\frac{\partial h_2}{\partial q_3}\right)=\rho f_3+\frac{1}{h_1h_2h_3}\left[\frac{\partial(h_2h_3 p_{13})}{\partial q_1}\right.$$

$$\left.+\frac{\partial(h_3h_1 p_{23})}{\partial q_2}+\frac{\partial(h_1h_2 p_{33})}{\partial q_3}\right]+\frac{p_{31}}{h_3h_1}\frac{\partial h_3}{\partial q_1}+\frac{p_{32}}{h_3h_2}\frac{\partial h_3}{\partial q_2}-\frac{p_{11}}{h_3h_1}\frac{\partial h_1}{\partial q_3}-\frac{p_{22}}{h_3h_2}\frac{\partial h_2}{\partial q_3},$$

where $d/dt=\partial/\partial t+h_1^{-1}v_1\partial/\partial q_1+h_2^{-1}v_2\partial/\partial q_2+h_3^{-1}v_3\partial/\partial q_3$, $p_{ij}$ are given in the first three equations, and in general $p_{ij}\neq p_{ji}$. By substituting $p_{ij}$ into the last three equations, one obtains the Navier-Stokes equations in this coordinate system. It should be noted that they may have different forms because derivatives of $h_1$, $h_2$, $h_3$ may satisfy some interrelating equations.